\documentclass[aps,prl,10pt,showpacs,
twocolumn,
groupedaddress]{revtex4-1}
\usepackage{graphicx,bm,wasysym,color}

\def\be{\begin{equation}} 
\def\ee{\end{equation}}  
\def\ba{\begin{eqnarray}}  
\def\ea{\end{eqnarray}}  
\def\bc{\begin{center}}  
\def\ec{\end{center}}

\begin{document}
\title{
Giant enhancement of the third harmonic in graphene integrated 
in a layered structure 
}

\author{N. A. Savostianova}
\affiliation{Institute of Physics, University of Augsburg, D-86135 Augsburg, Germany}
\author{S. A. Mikhailov}
\email[Email: ]{sergey.mikhailov@physik.uni-augsburg.de}
\affiliation{Institute of Physics, University of Augsburg, D-86135 Augsburg, Germany}

\date{\today}

\begin{abstract}
Graphene was shown to have strongly nonlinear electrodynamic properties. In particular, being irradiated by an electromagnetic wave with the frequency $\omega$, it can efficiently generate higher frequency harmonics. Here we predict that in a specially designed structure ``graphene -- dielectric -- metal'' the third-harmonic ($3\omega$) intensity can be increased by more than two orders of magnitude as compared to an isolated graphene layer.  
\end{abstract}

\pacs{78.67.Wj, 42.65.Ky, 73.50.Fq}

\maketitle

It was theoretically predicted \cite{Mikhailov07e} that, due to the ``ultra-relativistic'', massless energy dispersion of graphene electrons, 
\be E_{\pm}(\bm k)=\pm\hbar v_F|{\bm k}|,
\label{spectr}
\ee 
it should demonstrate a strongly nonlinear electrodynamic response; here $v_F\approx 10^8$ cm/s is the Fermi velocity of graphene and ${\bm k}$ is the electron wave-vector. Physically, this is due to the absence of inertia of graphene electrons: According to (\ref{spectr}), electrons can move only with the velocity $v_F$ in any directions, therefore, being placed in the oscillating external electric field they have to \textit{``instantaneously''} change their velocity from $+v_F$ to $-v_F$ in the return points. This leads to the emission of radiation at higher (multiple) frequencies as well as to other nonlinear phenomena. The efficiency of the nonlinear effects in graphene was predicted to be many orders of magnitude larger than in many other nonlinear materials \cite{Mikhailov07e}.

Experimentally, a strong nonlinearity of the graphene response has been confirmed at microwave \cite{Dragoman10} and optical \cite{Hendry10} frequencies. Further experimental studies of different nonlinear electrodynamic effects in graphene can be found in Refs. \cite{Dean10,Wu11,Zhang12,Bykov12,Kumar13,Hong13,An14,Lin14}. A quasi-classical theory of the nonlinear electrodynamic response of graphene, which is valid at relatively low (microwave, terahertz) frequencies $\hbar\omega\ll 2E_F$, has been developed in Refs. \cite{Mikhailov07e,Mikhailov08a,Mikhailov09a,Mikhailov11b,Mikhailov12b,Smirnova14,Yao14,Peres14}; here $E_F$ is the Fermi energy. This theory takes into account only the intra-band electronic transitions and ignores the inter-band ones. More general, quantum theories, which take into account both contributions, have been recently proposed in Refs. \cite{Cheng14a,Mikhailov14c,Cheng14b,Cheng15,MikhailovPreprint2015,Semnani15}. It was shown that, apart from a strong resonance at low ($\hbar\omega\ll 2E_F$) frequencies, the third-order nonlinear conductivity $\sigma_{\alpha\beta\gamma\delta}(\omega_1,\omega_2,\omega_3)$ demonstrates a number of resonances at the frequencies corresponding to the one-, two- and three-photon inter-band absorption. 

In Refs. \cite{Cheng14a,Mikhailov14c,Cheng14b,Cheng15,MikhailovPreprint2015,Semnani15} the third-order nonlinear response functions of graphene have been calculated for a single, freely hanging in vacuum (or in air) mono-atomic graphene layer. In reality graphene lies of a dielectric substrate (of thickness $d$). In this Letter we study the influence of the dielectric environment on the efficiency of the third harmonic generation and show that, depending on the ratio $d/\lambda_\omega$, as well as on physical properties of layers supporting graphene, the output third harmonic intensity can be both several orders of magnitude smaller and several orders of magnitude larger than in the isolated graphene layer. A proper choice of the geometrical parameters and physical properties of the dielectric environment is thus vitally important for the successful operation of graphene based nonlinear devices.

\begin{figure}
\includegraphics[width=8.4cm]{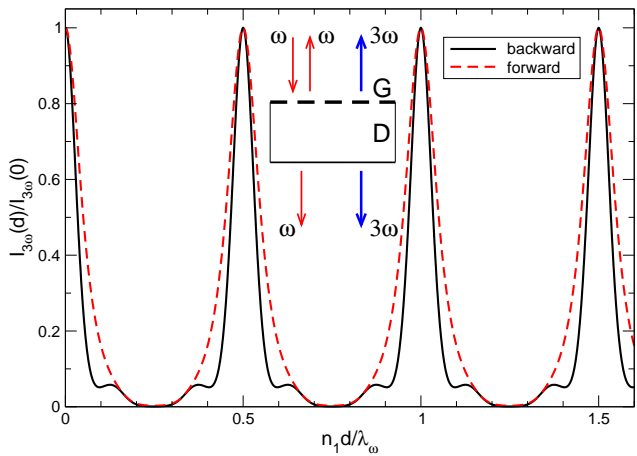}
\caption{(Color online) The normalized third harmonic intensity $I_{3\omega}(d)/I_{3\omega}(0)$ as a function of the dimensionless dielectric thickness $n_1d/\lambda_\omega$ for graphene lying on a dielectric slab. Here $I_{3\omega}(0)$ is the intensity of the $3\omega$-wave emitted from the isolated graphene layer in one (forward or backward) direction, $\lambda_\omega=2\pi c/\omega$ is the wavelength of radiation in vacuum and $n_1$ is the refractive index of the dielectric. The black solid (red dashed) curve shows the intensity of the third harmonic emitted in the backward (forward) direction. Inset shows the geometry of the described experiment; G = graphene, D = dielectric. }
\label{fig:diel}
\end{figure}

\begin{figure}
\includegraphics[width=8.4cm]{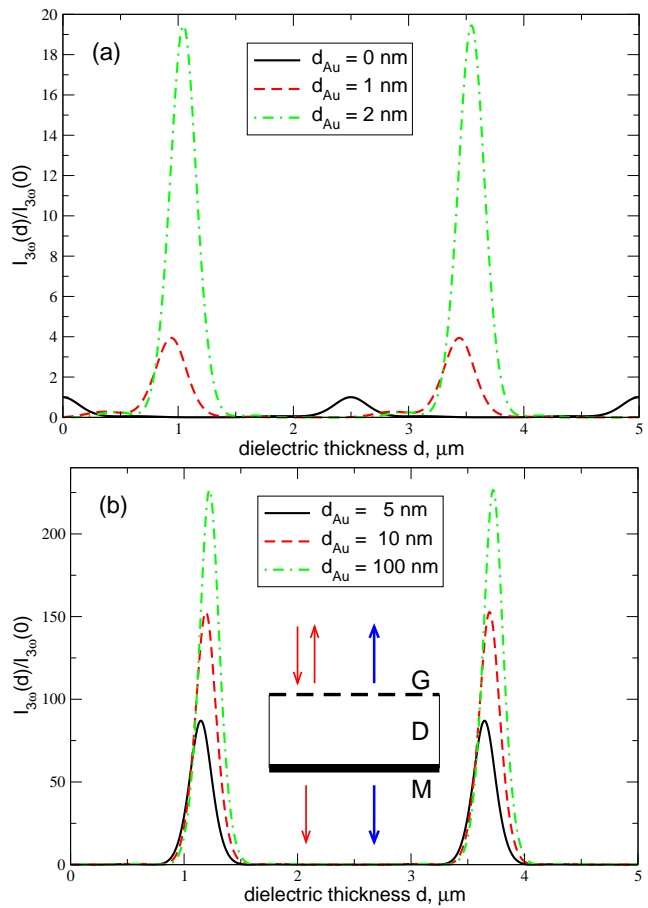}
\caption{(Color online) The third harmonic intensity emitted in the \textit{backward} (reflection) direction as a function of the dielectric thickness $d$, at different metal thicknesses $d_{\mathrm {Au}}$. Inset in (b) shows the geometry of the described experiment; G = graphene, D = dielectric, M = metal. The wavelength of the incident radiation is $\lambda_\omega=10$ $\mu$m.  }
\label{fig:metBackward}
\end{figure}

As an example, we consider a layered system ``graphene -- medium 1 -- medium 2'' where a two-dimensional graphene layer lies at the plane $z=0$ and each of the media is characterized by a thickness $d_j$ and a complex dielectric permittivity $\epsilon_j(\omega)$, $j=1,2$. We assume that a linearly polarized (in the $x$-direction) electromagnetic wave with the frequency $\omega$ and the power density $I_\omega$ is normally incident on the structure from the graphene side, see insets to Figs. \ref{fig:diel} and \ref{fig:metBackward}(b). The distribution of electromagnetic fields in the system is described by Maxwell equations with the dielectric function 
$\epsilon(\omega,z)$, being equal to $\epsilon_j(\omega)$ inside the $j$-th layer, and with the term $ j_\alpha(t)\delta(z)=\delta_{\alpha x}j(t)\delta(z)$, describing the current in the two-dimensional graphene layer. The current $j(t)=j^{(1)}(t)+j^{(3)}(t)$ in graphene has the linear and the third-harmonic components, $j^{(1)}(t)=\sigma_{xx}^{(1)}(\omega) E_{\omega}e^{-i\omega  t}$ and $j^{(3)}(t)=\sigma_{xxxx}^{(3)}(\omega,\omega,\omega)
E^3_{\omega}e^{-i3\omega  t}$,  
where $E_{\omega}\equiv E_{\omega,z=0}$ is the 
Fourier component of the self-consistent electric field at the plane $z=0$, and $\sigma^{(1)}$ and $\sigma^{(3)}$ are the first- and third-order conductivities of an isolated graphene layer. 

We solve the outlined nonlinear electrodynamic problem in two steps. First, we calculate, within the first-order response theory, Refs. \cite{Falkovsky07a,Gusynin07b,Mikhailov07d}, the $\omega$ Fourier components of the electric field and current, and relate the amplitude of the ac electric field $E_{\omega,z=0}$ at the plane $z=0$ to the amplitude of the incident wave $E_0$. Then we substitute the third-order current $j_{3\omega}=\sigma_{xxxx}^{(3)}(\omega,\omega,\omega)
E^3_{\omega}$ in Maxwell equations and calculate the amplitudes of the $3\omega$ Fourier components of the waves emitted in the forward and backward directions. We neglect the influence of the nonlinear effects on the amplitudes of $E_{\omega,z=0}$; these effects, determined by $\sigma_{xxxx}^{(3)}(\omega,\omega,-\omega)$, give small corrections to our results. Having found the amplitudes of the $3\omega$ components of the electric and magnetic fields, we calculate the intensity $I_{3\omega}$ of the third harmonic, as a function of the input-wave frequency $\omega$ and the thicknesses and material parameters of the media 1 and 2, and compare it with the third-harmonic intensity emitted by the isolated graphene layer.

\begin{figure}
\includegraphics[width=8.4cm]{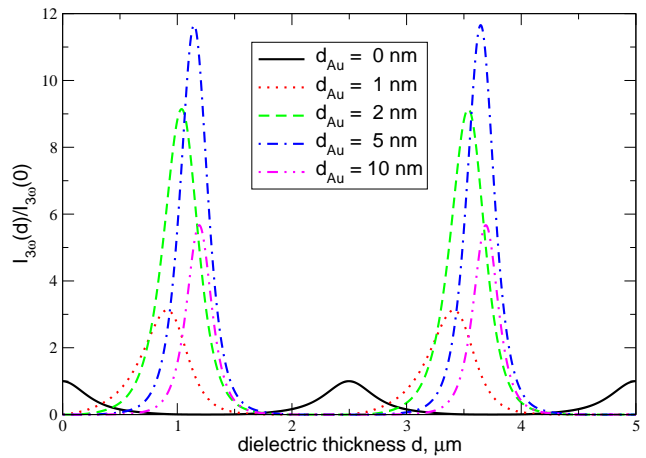}
\caption{(Color online) The normalized third harmonic intensity $I_{3\omega}(d)/I_{3\omega}(0)$, emitted in the \textit{forward} (transmission) direction, as a function of the dielectric thickness $d$, at different metal thicknesses $d_{\mathrm {Au}}$, in the structure shown in the inset to Fig. \ref{fig:metBackward}(b). The wavelength of the incident radiation is $\lambda_\omega=10$ $\mu$m. }
\label{fig:metForward}
\end{figure}

Figures \ref{fig:diel} -- \ref{fig:metBackward} show our results. We assume that the graphene layer has the electron density $n_s\simeq 3\times 10^{11}$ cm$^{-2}$ and the effective scattering time $\tau\simeq 1$ ps. The medium 1 is assumed to be a dielectric with the frequency independent refractive index $n_1=\sqrt{\epsilon_1}$ (a reasonable assumption if $\hbar\omega$ is much smaller than the band gap in the dielectric). The medium 2 (if present) is assumed to be a metal (gold) with the complex dielectric function $\epsilon_2(\omega)$ described by the Drude model with the plasma frequency $8.5$ eV and the scattering time $14$ fs. These numbers are taken from Ref. \cite{Olmon12} where it was shown that the Drude model gives a good description of the dielectric properties of gold at the photon energies lower than $\simeq 1$ eV. In Figures \ref{fig:metForward} -- \ref{fig:metBackward} we assume that the wavelength of the incident wave is $\lambda_\omega=10$ $\mu$m (the frequency $f=30$ THz, $\hbar\omega=0.12$ eV). 

Figure \ref{fig:diel} shows the third harmonic intensity emitted from graphene lying on the dielectric substrate with the thickness $d$, as a function of the dimensionless dielectric thickness $n_1d/\lambda_\omega$. The intensity $I_{3\omega}$ strongly oscillates as a function of $d$, with maxima corresponding to the integer ($m$) numbers of half-wavelengths ($\lambda_\omega/n_1$) inside the dielectric slab,
\be 
d=\frac{\lambda_\omega}{2n_1}m,\ \ \ m=0,1,2\dots;\label{optimal-thickness}
\ee 
under these conditions the field $E_\omega$ at $z=0$ is maximal. In the maxima the values of $I_{3\omega}(d)$ are the same as in the isolated graphene $I_{3\omega}(0)$, but in the minima they can be several orders of magnitude smaller [$I_{3\omega}(d)/I_{3\omega}(0)\lesssim  10^{-3}$ in Fig. \ref{fig:diel}; here $I_{3\omega}(0)$ is the third-harmonic intensity emitted only in one, forward or backward, direction]. A proper choice of the dielectric thickness is thus of extreme importance for observation of the third harmonic generation. In Fig. \ref{fig:diel} one also sees a difference in the intensity of the $3\omega$-waves emitted in the forward and backward directions: the intensity of the backward radiation is smaller and has additional resonances  corresponding to the interference condition for the third harmonic, $d\simeq (\lambda_{3\omega}/2n_1)m$, $m=0,1,2,\dots$. 

Now consider the case when a thin metallic (Au) layer with the thickness $d_{\mathrm {Au}}$ covers the backside of the dielectric substrate. In this case the third harmonic intensity strongly depends on whether the $3\omega$-wave is emitted in the forward (transmission) or in the backward (reflection) direction. Figure \ref{fig:metForward} shows the normalized intensity of the third harmonic, emitted by the structure in the \textit{forward} direction, as a function of the dielectric thickness $d$ at several different metal thicknesses $d_{\mathrm {Au}}$. 

A few interesting features are seen in Fig. \ref{fig:metForward}. First, when the metal thickness $d_{\mathrm {Au}}$ increases from $d_{\mathrm {Au}}=0$ nm (the black solid curve; corresponds to the red dashed curve in Fig. \ref{fig:diel}) up to $d_{\mathrm {Au}}\simeq 5$ nm, the maxima of $I_{3\omega}$ tend to new positions determined by the relation 
\be 
d=\frac{\lambda_\omega}{2n_1}\left(m+\frac 12\right),\ \ \ m=0,1,2\dots,\label{optimal-thickness-metal}
\ee 
and the intensity $I_{3\omega}$ of the transmitted third-harmonic signal substantially (by more than one order of magnitude) increases. The quarter-wavelength shift in (\ref{optimal-thickness-metal}), as compared to (\ref{optimal-thickness}), results from the boundary condition for the tangential electric field at the metallic plane $z=d$, $E_{\omega,z=d}\simeq 0$. This also leads to a larger fundamental-frequency electric field $E_{\omega,z=0}$ at the graphene plane $z=0$, and hence, to a larger third-harmonic signal. 

When $d_{\mathrm {Au}}$ grows further, the intensity $I_{3\omega}$ reaches its maximum and then decreases (at $d_{\mathrm {Au}}\gtrsim 5$ nm), since at large $d_{\mathrm {Au}}$ the metallic layer becomes opaque both for the fundamental ($\omega$) and for the third ($3\omega$) harmonic. 

Figure \ref{fig:metBackward} shows the intensity of the third harmonic emitted in the \textit{backward} direction. One sees that the maxima of the curves $I_{3\omega}(d)$ also move to the $d$-values (\ref{optimal-thickness-metal}) when $d_{\mathrm {Au}}$ increases. However, in contrast to the forward-direction emission, in these $d$-points $I_{3\omega}$ \textit{monotonously grows} with $d_{\mathrm {Au}}$ and \textit{saturates} at $d_{\mathrm {Au}}\gtrsim 100$ nm, reaching the value $I_{3\omega}(d)/I_{3\omega}(0)\simeq 226$. If to take into account that the \textit{total} third-harmonic intensity emitted by an isolated graphene layer in both (forward and backward) directions equals $2I_{3\omega}(0)$, one sees that just the metalization of the backside of the substrate may increase the total emitted $3\omega$ power by more than two orders of magnitude (by the factor of $\simeq 113$). 

Such a giant enhancement of the third harmonic intensity is explained by the interference of both the incident wave ($\omega$) and its third harmonic ($3\omega$) in the dielectric slab. Indeed, in the case of a single (isolated) graphene layer in air the third order current in the layer is of order $j_{3\omega}\simeq \sigma^{(3)}E_{\omega,z=0}^3\approx \sigma^{(3)}E_{0}^3$, where $E_0$ is the amplitude of the incident wave (in this simple estimate we ignore the difference between $E_{\omega,z=0}$ and $E_{0}$). The electric and magnetic fields of the emitted third harmonic are proportional to $j_{3\omega}$, $E_{3\omega}\propto H_{3\omega}\propto \sigma^{(3)}E_{0}^3$, and the intensity of the emitted radiation (in one direction) is proportional to $|E_{3\omega}|^2$. The total emitted power (both in the backward and forward directions) is then $I_{3\omega}^{\mathrm{single\ layer}}\simeq 2|\sigma^{(3)}E_0^3|^2$.

If the graphene layer lies on the dielectric slab metalized on the back side, the fundamental-frequency electric field at the plane $z=0$ is twice as large (in the interference maximum) as the incident wave field, $E_{\omega,z=0}\approx 2E_0$, due to the reflection of the wave from the back-side mirror. The third-order current is then $j_{3\omega}\simeq \sigma^{(3)}E_{\omega,z=0}^3\approx 2^3 \sigma^{(3)}E_{0}^3$. The fields $E_{3\omega}$, $H_{3\omega}$ of the excited third-harmonic wave, emitted from the graphene plane $z=0$ in both directions, are proportional to the current $j_{3\omega}$, but since the $3\omega$-wave is reflected from the mirror again, they should be multiplied by two once more, $E_{3\omega}\propto H_{3\omega}\propto 2j_{3\omega}\propto 2^4 \sigma^{(3)}E_{0}^3$. The total emitted power (only in the backward direction) is then proportional to $I_{3\omega}^{\mathrm{slab}}\simeq 2^8 |\sigma^{(3)}E_0^3|^2$. Comparing $I_{3\omega}^{\mathrm{slab}}$ with $I_{3\omega}^{\mathrm{single\ layer}}$ we see that the amplification factor, due to the substrate with a metalized back side, is about $2^7=128$. The slightly smaller factor $\simeq 113$ obtained in the more accurate theory above (Fig. \ref{fig:metBackward}) is due to the small difference between $E_{\omega,z=0}$ and $E_{0}$ which we have neglected. 

Our results thus show that, placing graphene on the surface of a dielectric slab with a metalized back side one can increase the generated third-harmonic intensity by more than two orders of magnitude (a further enhancement of the third harmonic can be achieved by placing graphene inside a Fabry-P\'erot-type cavity with a high quality factor). Parameters of such a graphene-based frequency multiplier should however be carefully chosen since the dielectric thickness $d$ and the input radiation frequency $\omega$ (or the wavelength $\lambda_\omega$) should be related by the resonant conditions (\ref{optimal-thickness-metal}). As seen from Figs. \ref{fig:diel} -- \ref{fig:metBackward}, deviations from these relations may \textit{suppress} the output $3\omega$ signal by orders of magnitude. 
In addition, a further (resonant) enhancement of the third harmonic at infrared frequencies can be achieved by tuning the density of electrons $n_s$ in graphene, since the positions of the inter-band resonances depend on the Fermi energy. 

It should be noticed that numerical values of the third-order parameters of graphene, experimentally measured in different papers, quite substantially differ from each other, see, e.g., a discussion in \cite{Cheng14a}. This may be due to a better or worse fulfillment of the resonant conditions (\ref{optimal-thickness}) or (\ref{optimal-thickness-metal}) in different experiments. 
It can also be noticed that another way to increase the third harmonic generation (by a factor of $\sim 5$) was proposed in Ref. \cite{Smirnova15}. In that paper the radiation propagates \textit{along} the layered structure, containing graphene, the low-frequency regime is considered ($\hbar\omega<2E_F$) and the third-harmonic enhancement is achieved by utilizing cascaded processes. A Fabry-P\'erot cavity with a graphene layer inside was proposed to be used for the enhancement of the linear absorption in intrinsic ($E_F=0$) graphene in Ref. \cite{Ferreira12}.

To summarize, we have studied the third-harmonic generation effect in the structure graphene--dielectric--metal and found the optimal operation conditions of such a frequency multiplier, Eq. (\ref{optimal-thickness-metal}). Properly choosing the system parameters one can get a giant, resonant enhancement of the up-conversion efficiency. Conversely, under the off-resonant conditions the device efficiency is suppressed by many orders of magnitude. The correct choice of the device parameters, Eq. (\ref{optimal-thickness-metal}), is thus of particular importance for its proper operation.

The work was supported by the European Union under the Program Graphene Flagship (No. CNECT-ICT-604391).


%


\end{document}